\documentclass[%
 aps,
 amsmath,amssymb,
 aps,
]{revtex4}
\usepackage{amssymb}
\usepackage{amsfonts}
\usepackage{amsmath}
\usepackage{epsfig, color}

\usepackage{graphicx}
\graphicspath{%
    {converted_graphics/}
    {/}
}
\usepackage{comment}
\usepackage{amsfonts,amssymb,stmaryrd,amsthm}
\usepackage{amsmath}
\usepackage{graphicx}
\usepackage{enumerate}
\usepackage{subfigure}
\usepackage{amscd}
\usepackage{pb-diagram}
\usepackage[toc]{appendix}
\usepackage{afterpage}
\usepackage{bm}

\begin{document}

\title{Traveling Wave Solutions to Fifth- and Seventh-order Korteweg-de Vries Equations: Sech and Cn Solutions}
	
\author{Stefan C.\ Mancas}
\email{mancass@erau.edu}
\affiliation{Department of Mathematics, Embry--Riddle Aeronautical University \\ Daytona Beach, FL 32114-3900, USA}

\author{Willy A.\ Hereman}
\email{whereman@mines.edu}
\affiliation{Department of Applied Mathematics and Statistics\\Colorado School of Mines, Golden CO 80401-1887, USA}

\begin{abstract}
In this paper we review the physical relevance of a Korteweg-de Vries (KdV) equation with higher-order
dispersion terms which is used in the applied sciences and engineering.
We also present exact traveling wave solutions to this generalized KdV equation using an
elliptic function method which can be readily applied to any scalar evolution or wave equation
with polynomial terms involving only odd derivatives.
We show that the generalized KdV equation still supports hump-shaped solitary waves
as well as cnoidal wave solutions provided that the coefficients satisfy specific algebraic constraints.

Analytical solutions in closed form serve as benchmarks for numerical solvers or comparison with experimental data.
They often  correspond to homoclinic orbits in the phase space and serve as separatrices of stable and unstable regions.
Some of the solutions presented in this paper correct, complement, and illustrate results previously
reported in the literature, while others are novel.
\end{abstract}

\medskip

\maketitle

\section{Introduction}
\label{intro}
Motivated by the work of Malfliet, Hereman, and collaborators \cite{Bald,Malf1,Malf2,Malf3}
we will apply an elliptic function method to a nonlinear dispersive seventh-order KdV equation which,
in general, takes the form
\begin{equation}
\label{7thKdVorg}
u_{t} + a uu_{x} + b u_{3x} + c u_{5x} + d u_{7x} = 0,
\end{equation}
with real constant coefficients, and where $u_{nx} = \frac{{\partial}^n u}{{\partial}x^n}$ for positive integer $n$.

The generalized Korteweg-de Vries (KdV) equation (\ref{7thKdVorg}) has been widely used in the applied sciences and engineering.
The case $c = d = 0$ corresponds to the ubiquitous KdV equation, \cite{KdV} describing, for example,
shallow water waves and ion-acoustic waves in plasmas.
In this paper we do not discuss this well-studied case.
The case $c \ne 0,\, d = 0$ was discussed by Hasimoto \cite{Has} for shallow water waves near some critical value
of surface tension, and also was investigated numerically in studies of magneto-acoustic as well as hydrodynamic waves
in a cold collision-free plasma by Kawahara \cite{Kawa5}, Kakutani and Ono \cite{Kaku}, respectively.
For completeness, we will present the cnoidal (cn) and sech solutions for the KdV equation with a fifth-order
dispersion term.
The paper focuses on (\ref{7thKdVorg}) when both higher-order dispersive terms are present
($c \ne 0, \, d \ne 0),$
for which some new results are presented.
The nature and relevance of the dispersive terms depends on the physical context or application.
Such applications include shallow water waves, electrical pulses in transmission lines, waves in plasmas, etc.

As illustrated by Craig {\it et al.} \cite{Craig} for Boussinesq and KdV equations,
the inclusion of the next higher order in the equations may turn an ill-posed problem into a well-posed one.
Pomeau {\it et al.} \cite{Pom} investigated if (\ref{7thKdVorg}), viewed as a perturbed KdV equation,
would still possess a well-localized solitary wave solution and how higher-order terms would
affect their stability.
They showed that the solution is no longer strictly localized but develops an infinite oscillating tail
which was already observed independently by Kawahara \cite{Kawa5} and experimentally verified by Nagashima \cite{Naga1}
in a transmission line with a large number of inductors and capacitors.
Eq.\ (\ref{7thKdVorg}) was also derived by Rosenau using a quasi-continuous formalism that included higher order
discrete effects \cite{Ros1,Ros2}.

Closed-form analytic solutions of (\ref{7thKdVorg}) were first obtained by Ma \cite{Ma} using a trial function method.
He found a solitary wave solution with a $\mathrm{sech}^6$-term which, unfortunately, does not satisfy the PDE
due to misprints as pointed out by Duffy and Parkes \cite{Duffy}.
Later, Parkes {\it et al.} \cite{Parkes} and Yang \cite{Yang} introduced recursive methods to find exact solutions
to KdV-type equations with any number of odd derivatives but the solutions were not written explicitly.
Many researchers \cite{Ros1,Ros2,Dai,Hunter83,Her85,Her86,Kud3,Manc}
have found solitary waves solutions to the fifth-order KdV equation
or analyzed their stability \cite{Andrade}.
In general, most of the solutions found were written in terms of hyperbolic or exponential functions,
with the exception of the cnoidal $({\rm cn})$ waves for the transmission line
(when $b = d = 0$ in (\ref{7thKdVorg}))
derived by Yamamoto and Takizawa, \cite{Yama7} and the periodic solutions
computed by Kano and Nakayama, \cite{Kano4} and Khater {\it et al.} \cite{Khater}
For the full fifth-order equation Huang {\it et al.} \cite{Huang}
found ${\rm sech}^4$-solutions but concluded incorrectly
there are no explicit solutions of type $D + A\,\mathrm{cn}^4(Bx-Ct; k^2)$ unless $k^2 = 1.$
In fact, the contrary is true for Parkes {\it et al.} \cite{Parkesbis} have found periodic solutions using
$\mathrm{sn}$- and $\mathrm{cn}$-function methods, and Mancas \cite{Manc} has obtained explicit solutions
when $k^2 = \frac{1}{2}$ using an elliptic function method.

Some of the solutions presented in this paper correct, complement, and illustrate results previously reported
in the literature.
To our knowledge, the general cn-solutions given in Sects.~\ref{cnsolfifthorderKdV} and~\ref{cnsolseventhorderKdV}
are new.

Solutions in closed analytical form serve as benchmarks for numerical solvers \cite{Kalischetal,Simpson}
and comparison with experimental data.
From a dynamical systems perspective, closed-form solutions play an important role in studies
of bifurcations of traveling wave solutions \cite{Longetal,Lietal}.
Usually, hump-type solutions correspond to homoclinic orbits, kink-type solutions to heteroclinic orbits
(also called connecting orbits), and periodic traveling waves to periodic orbits in phase space.
When available, such exact solutions describe the separatrices which separate stable and unstable
regions in the phase space.
Finally, exact solutions evaluated at $t=0$ have been used as the initial conditions $u(x,0)$ to test various
perturbation methods \cite{Arora,Din,Dinbis,Din2bis} applied to (\ref{7thKdVorg}).

Eq.\ (\ref{7thKdVorg}) can be written in Hamiltonian form \cite{Drazin} as
\begin{equation}
\label{HamilKdV}
u_t = \frac{\partial}{\partial x} \left(\frac{\delta \mathcal H}{\delta u}\right),
\end{equation}
where $\frac{\delta}{\delta u}$ is the variational derivative and
\begin{equation}
\label{Hamiltonian}
\mathcal {H} = - \frac{a}{6} u^3 + \frac{b}{2} {u_x}^2 -\frac{c}{2} {u_{2x}}^2 + \frac{d}{2} {u_{3x}}^2
\end{equation}
is the Hamiltonian.
The Hamiltonian expresses the conserved energy density because the energy integral
$\mathcal I = \int_{-\infty}^{\infty} \mathcal {H} dx$ does not change with time.

Eq.\ (\ref{7thKdVorg}) can be re-scaled using the transformations
$u(x,t) = A U(X,T)$ with $X = B x$, $T = C t$.
For $A = - \frac{b^2}{ac}$, $B = \epsilon \sqrt{-\frac{b}{c}}$,
$C = -\frac{b^2 \epsilon}{c} \sqrt{-\frac{b}{c}}$, $\epsilon = \pm 1$ with $bc < 0,$
one obtains the non-dimensional equation
\begin{equation}
\label{7thKdVXT}
U_{T} + U U_{X} + U_{3X} - U_{5X} + \alpha U_{7X} = 0,
\end{equation}
with  $\alpha=\frac{bd}{c^2}.$
The fifth-order KdV equation corresponds to the case $d=0$, which implies $\alpha=0$.
To find traveling wave solutions to (\ref{7thKdVXT}) we use the ansatz $U(\xi) = U(X-\text{v}T)$,
where $\text{v}$ is the velocity of a unidirectional wave traveling in the positive $X$ direction.
After one integration with respect to $\xi$ one obtains a first integral,
\begin{equation}
\label{conslaw1}
\text{v} U - \frac{1}{2} U^2 - U_{2\xi} + U_{4 \xi} - \alpha U_{6\xi} = {\mathcal{C}},
\end{equation}
where ${\mathcal{C}}$ is an integration constant.
Eq.\ (\ref{conslaw1}) is a conservation law in the traveling frame of reference.
The first term is proportional to the mass density; the remaining terms express the mass flux.
After multiplication of (\ref{conslaw1}) with $U_{\xi}$ and integration one gets
\begin{equation}
\label{conslaw2short}
\frac{\text{v}}{2} U^2 - \frac{1}{6} U^3 - \frac{1}{2} {U_{\xi}}^2 - \frac{1}{2} {U_{2\xi}}^2 + U_\xi U_{3 \xi}
- \alpha \left( \frac{1}{2} {U_{3\xi}}^2 - U_{2\xi} U_{4\xi} + U_{\xi} U_{5\xi} \right) - {\mathcal{C}} U
+ {\mathcal{D}} = 0,
\end{equation}
which, upon substitution of ${\mathcal{C}}$ from (\ref{conslaw1}), leads to a second conservation law:
\begin{equation}
\label{conslaw2long}
\frac{\text{v}}{2} U^2 -\! \frac{1}{3} U^3 +\! \frac{1}{2} {U_{\xi}}^2 -\! U_\xi U_{3 \xi}
  +\! \frac{1}{2} {U_{2\xi}}^2 -\! U U_{2\xi} +\! U U_{4\xi}
+ \! \alpha \left( U_\xi U_{5\xi} -\! U_{2\xi} U_{4\xi} +\! \frac{1}{2} {U_{3\xi}}^2 -\! U U_{6\xi} \right)
=  {\mathcal{D}},
\end{equation}
where ${\mathcal{D}}$ is also a constant of integration.
The first term is proportional to the momentum density; the remaining terms express the momentum flux.
%
%
\section{Elliptic Function Method}
\label{ellipticmethod}
To seek analytic solutions of (\ref{conslaw1}) we use an elliptic function method which
can be readily applied to any scalar evolution or wave equation with polynomial terms
involving only odd derivatives.
We assume that $U$ has a polynomial expansion of the form
\begin{equation}
\label{UEFM}
U(\xi) = \sum_{i=0}^N A_i Y(\xi)^i,
\end{equation}
with constant coefficients $A_i,$ and where $Y(\xi)$ satisfies the equation
\begin{equation}
\label{YxiEFM}
{Y_\xi}^2 = \sum_{i=0}^3 a_i Y(\xi)^i \equiv Q_3(Y).
\end{equation}
To determine the highest exponent $(N)$, we substitute $U(\xi) = Y(\xi)^N $ into (\ref{conslaw1}),
with the even derivatives given by $Y_{n \xi} = Y^{\frac{n+2}{2}}$ from (\ref{YxiEFM}),
and equate the highest powers of $Y$.
They occur in the highest order term ($U_{6\xi}$ for the seventh-order equation
or $U_{4\xi}$ for the fifth-order equation), and the nonlinear term $(U^2)$.
Doing so, $Y^{N+3} = Y^{2N}$ yields $N=3$ (seventh-order case)
and $Y^{N+2} = Y^{2N}$ yields $N=2$ (fifth-order case).
To cover both cases at once, we set $N=3$ in the computations below.
Based on (\ref{UEFM}), it suffices to set $A_3 = \alpha = 0$ to get the results for $N=2.$
%
%
\section{Sech Solutions}
\label{sechsolutions}
To obtain solitary wave solutions we require that $Y=0$ is a double root of $Q_3(Y).$
This condition is achieved by choosing zero boundary conditions,
$Y, Y_{\xi}, Y_{2\xi}\rightarrow 0$ as $|\xi|\rightarrow \infty$.
These conditions are met when $a_0 = a_1 = 0$ in (\ref{YxiEFM}).
Hence,
\begin{equation}
\label{Yxisech}
{Y_\xi}^2 = a_2 Y^2 + a_3 Y^3 = Y^2 (a_2 + a_3 Y),
\end{equation}
with solitary wave solution
\begin{equation}
\label{Ysechsq}
Y(\xi) = - \frac{a_2}{a_3}\,\mathrm{sech}^2\left[\frac{1}{2} \sqrt{a_2}(\xi - \xi_0)\right],
\quad a_2 > 0, \; a_3 \ne 0,
\end{equation}
where $\xi_0$ is an arbitrary constant.
Now, (\ref{UEFM}) must be substituted into (\ref{conslaw1}).
To do so, we first differentiate $U(\xi)$ from (\ref{UEFM}) with respect to $\xi$ and
substitute ${Y_\xi}^2$ from (\ref{Yxisech}) together with all higher order derivatives,
$Y_{2\xi}$ through $Y_{6\xi},$ to express all the even derivatives of $U$ as polynomials in $Y$.
The explicit expressions of the derivatives of $U$ are given in the Appendix.
Next, we substitute the derivatives (\ref{app:sechUderiv}) into (\ref{conslaw1})
to obtain the sixth-degree polynomial
$ \sum_{i=0}^6 r_i Y(\xi)^i, $
which must vanish identically.
The expressions for the coefficients $r_i$ are given in (\ref{app:sechrcoef}).
The equations, $r_i = 0,$ must be solved for the $A_i$ in terms of the $a_i, \alpha$ and $\text{v}$,
some of which become constrained.
The integration constants $\mathcal{C}$ and $\mathcal{D}$  typically depend on $\text{v}$.
\subsection{Sech solutions for the fifth-order KdV equation}
\label{sechfifthKdV}
To compute sech solutions for (\ref{7thKdVXT}) with $\alpha = 0$, we set $A_3 = 0$
since $N=2$ in (\ref{UEFM}) and discard $r_5$ and $r_6$.
With regard to (\ref{app:sechrcoef}), solving $r_2 = r_3 = r_4 = 0$ gives
\begin{equation}
\label{cAsech5thKdV}
A_0 = \text{v} - \frac{31}{507} - \frac{7}{507} p (10 + p), \;
A_1 = -\frac{70}{13} a_3 (1 - p), \;
A_2 = 105 {a_3}^2,
\end{equation}
where $p = 13 a_2$.

Solving $r_0 = r_1 = 0,$ after substituting these coefficients, yields
\begin{equation}
\label{pCsech5thKdV}
p = 1, \;\;
\mathcal {C} = \frac{\text{v}^2}{2} - \frac{2^3\, 3^4}{13^4}.
\end{equation}
From (\ref{pCsech5thKdV}), $A_0=\text{v}-\frac{36}{169}$ and $A_1$ = 0.
According to (\ref{Ysechsq}), we then get
\begin{equation}
\label{Usechquart5thKdV}
U(\xi) =
\text{v}
-\frac{36}{169} +\frac{105}{169}\,\mathrm{sech}^4\left[\frac{1}{2 \sqrt{13}}(\xi-\xi_0)\right],
\end{equation}
where $\text{v}$ is arbitrary.

\begin{figure}[hb!]
\centering
\includegraphics[height=1.5in]{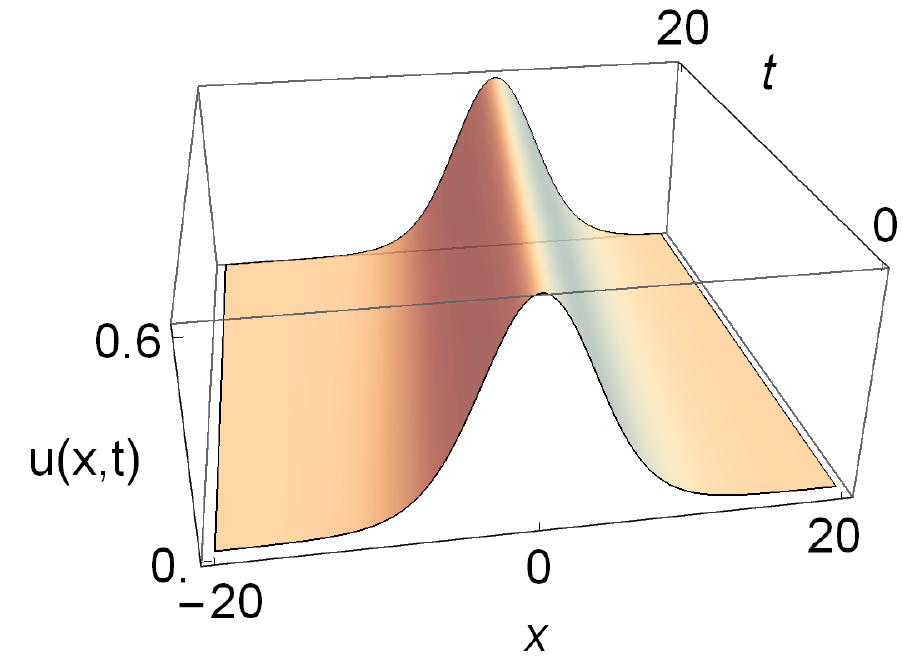}
\includegraphics[height=1.5in]{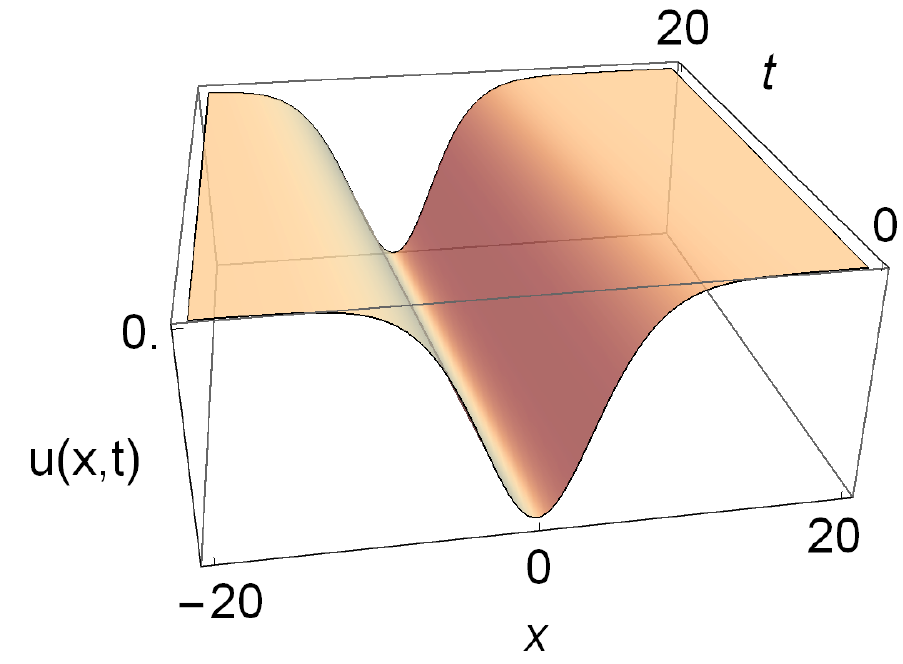}
\caption{
Solution (\ref{usechquart5thKdVspec}) for
(a)\, $a=b=-c=1$ (hump), and
(b)\, $a=\frac{9}{10}$, $b=-\frac{11}{10}$, $c=\frac{41}{40}$ (dip).}
\label{fig:sech-5thkdv-brightdark}
\end{figure}

Replacing $\xi$ by $X-\text{v} T$ and converting (\ref{Usechquart5thKdV}) into the original variables yields
\begin{equation}
\label{usechquart5thKdVorg}
u(x,t) =
\frac{b^2}{ac} \left\{\frac{36}{169}-\text{v} - \frac{105}{169}\,\mathrm{sech}^4\left[\frac{1}{2 \sqrt{13}}
\left(\delta_1+\sqrt{-\frac{b}{c}}(x+\frac{b^2\text{v}}{c} t)\right)\right] \right\},
\end{equation}
which satisfies (\ref{7thKdVorg}) when $d =0$.
This solution represents a family of solitary waves that travel to the right when
$\frac{\text{v}}{c} < 0 $, and to the left otherwise, while shifting up or down
on the vertical axis as $\text{v}$ changes.
For an unshifted wave $(A_0 = 0)$, one gets $\text{v} = \frac{36}{169} \approx 0.2130.$
Setting $\delta_1 = 0$, (\ref{usechquart5thKdVorg}) then simplifies into
\begin{equation}
\label{usechquart5thKdVspec}
u(x,t) =
-\frac{105 \,b^2}{169 \,ac}\,\mathrm{sech}^4\left[\frac{1}{2 \sqrt{13}}
\left(\sqrt{-\frac{b}{c}}(x+\frac{36 \, b^2}{169 \, c} t)\right)\right],
\end{equation}
which appeared earlier in the literature \cite{Hunter83,Her85,Her86}.
Fig.\ \ref{fig:sech-5thkdv-brightdark} illustrates solution (\ref{usechquart5thKdVspec})
for $a = b = - c = 1,$ and $a = \frac{9}{10}$, $b=-\frac{11}{10}$, and $c=\frac{41}{40},$
resulting in a hump and a dip, respectively.
%
%
\subsection{Sech solutions for the seventh-order KdV equation}
\label{sechseventhKdV}
To find sech solutions for (\ref{7thKdVXT}) using the expressions in (\ref{app:sechrcoef}),
we first solve $r_3 = r_4 = r_5 = r_6 = 0$, yielding
\begin{align}
\label{cAsech7thKdV}
& A_0 = \text{v} - \frac{231}{50} {a_2}^2 -22 \alpha {a_2}^3
  + \frac{(334)7^2 - (681)5^3 \alpha}{(262)5^5 \alpha^2} + \frac{231 (49 - 500 \alpha) a_2}{(262)5^3 \alpha},
\nonumber \\
& A_1 = \frac{231 (3 a_3)}{(262)5^3 \alpha} \left[49-50 \alpha \left(10-131 a_2 \right)-(524)5^3 \alpha^2 {a_2}^2 \right],
\\
& A_2 = \frac{231}{10}(3 a_3)^2 \left(1 - 50 \alpha a_2 \right), \;\;
A_3 = - 385 \alpha (3 a_3)^3.
\nonumber
\end{align}
After substituting these coefficients into $r_1 = r_2 = 0,$ we must distinguish between two cases:
$a_2^{+}=\frac{100}{2159},\,\alpha^{+}=\frac{2159}{10^4}$
and
$a_2^{-}=\frac{50}{769},\,\alpha^{-} =\frac{769}{2500}$.
Furthermore, $r_0 = 0$ determines the integration constant
$\mathcal{C} = \frac{1}{2} A_0 \left( 2 \text{v} - A_0 \right)$
which for the respective cases becomes
$\mathcal{C}^{+} = \frac{\text{v}^2}{2} - \frac{(5)71^2\,10^7}{17^4\,127^4}$
and
$\mathcal{C}^{-} = \frac{\text{v}^2}{2} - \frac{(2) 3^4\,10^8}{769^4}.$
\subsubsection{Case $(a_2^{+}, \,\alpha^{+})$}
\label{case1sechseventhKdV}
In this case, system (\ref{cAsech7thKdV}) gives
\begin{equation}
\label{cAsech7thKdVPlus}
A_0 = \text{v} -\frac{(71) 10^4}{17^2 127^2}, \;
A_1 = 0, \,
A_2 = - \frac{231}{20} (3 a_3)^2, \, \mathrm{and} \;\;
A_3 = -\frac{(17)(77)(127)}{(2) 10^3} (3 a_3)^3.
\end{equation}
Substituting (\ref{Ysechsq}) into (\ref{UEFM}) yields
\begin{equation}
\label{Usechsix7thKdVPlus}
U(\xi) = \text{v} - \frac{(71) 10^4}{17^2 127^2} + \frac{(385) 3^3 10^2}{17^2 127^2}
\,\mathrm{sech}^4\eta \,(1 + \mathrm{sech}^2 \eta),
\end{equation}
where $\eta=\frac{5(\xi-\xi_0)}{\sqrt{2159}}$.
Eq.\ (\ref{Usechsix7thKdVPlus}), in which $\text{v}$ and $\xi_0$ are arbitrary constants,
solves (\ref{conslaw1}) provided $\alpha = \frac{2159}{10^4}$ and $\mathcal C=\mathcal C^{+}$.
Converting (\ref{Usechsix7thKdVPlus}) into the original variables yields
\begin{eqnarray}
\label{usechsix7thKdVPlusorg}
u(x,t) = &
\frac{b^2}{17^2\,127^2 ac} \Bigg\{
(71)10^4-17^2 127^2\text{v}-(385) 3^3 10^2\,\mathrm{sech}^4\left[\frac{5}{\sqrt{2159}}
\left(\delta_1+\sqrt{-\frac{b}{c}}(x+\frac{b^2\text{v}}{c}t)\right)\right]
\nonumber \\
& \left\{ 1 + \mathrm{sech}^2\left[\frac{5}{\sqrt{2159}}
\left(\delta_1+\sqrt{-\frac{b}{c}}(x+\frac{b^2\text{v}}{c} t)\right)\right] \right\} \Bigg\},
\end{eqnarray}
where both the velocity $\text{v}$ and phase constant $\delta_1$ are arbitrary.
Solution (\ref{usechsix7thKdVPlusorg}) satisfies (\ref{7thKdVorg}) provided
$d = \frac{2159\,c^2}{10^4\,b}.$
When $A_0 = 0$ the wave speed becomes $\text{v}^{+} = \frac{(71)\,10^4}{17^2\,127^2} \approx 0.1523$
and $\mathcal{C}^{+} = 0.$
Physically, this means that the solitary wave solution is not shifted on the vertical axis but tends
to zero as $|x| \rightarrow \infty$.
Furthermore, for $\delta_1=0$ and $\text{v}$ replaced by $\text{v}^{+},$ (\ref{usechsix7thKdVPlusorg})
simplifies into
\begin{eqnarray}
\label{usechsix7thKdVPlusspec}
u(x,t) =
& -\, \frac{(385) 3^3 10^2 \,b^2}{17^2\,127^2 ac}\,\mathrm{sech}^4 \left[\frac{5}{\sqrt{2159}}
\sqrt{-\frac{b}{c}} \left(x + \frac{(71)\,10^4\,b^2}{17^2\,127^2\,c}t \right)\right]
\nonumber \\
& \left\{1 + \mathrm{sech}^2 \left[\frac{5}{\sqrt{2159}}
\sqrt{-\frac{b}{c}} \left( x +\frac{(71)\,10^4\,b^2}{17^2\,127^2\,c} t \right)\right] \right\}.
\end{eqnarray}
\subsubsection{Case $(a_2^{-}, \,\alpha^{-})$}
\label{case2sechseventhKdV}
For this case, system (\ref{cAsech7thKdV}) becomes
\begin{equation}
\label{cAsech7thKdVMinus}
A_0 = \text{v} - \frac{(18)\,10^4}{769^2}, \;\;
A_1 = 0, \;\;
A_2 = 0, \;\; \mathrm{and} \;\;
A_3 = -\frac{(77)\,(769)}{500}(3 a_3)^3.
\end{equation}
Upon substitution of (\ref{Ysechsq}) into (\ref{UEFM}), one gets
\begin{equation}
\label{Usechsix7thKdVMinus}
U(\xi) = \text{v} - \frac{(18)\,10^4}{769^2} + \frac{(154)\,3^3\,5^3}{769^2}\,\mathrm{sech}^6 \chi,
\end{equation}
where $\chi = \frac{5(\xi - \xi_0)}{\sqrt{1538}}$.
Eq.\ (\ref{Usechsix7thKdVMinus}), in which $\text{v}$ and $\xi_0$ are arbitrary constants,
solves (\ref{conslaw1}) provided $\alpha = \frac{769}{2500}$ and $\mathcal C = \mathcal C^{-}$.
In the original variables (\ref{Usechsix7thKdVMinus}) reads
\begin{equation}
\label{usechsix7thKdVMinusorg}
u(x,t) = \frac{b^2}{769^2 ac}
\left\{(18)10^4 -\! 769^2 \text{v} -\!(154) 3^3 5^3
\,\mathrm{sech}^6 \left[\!\frac{5}{\sqrt{1538}}
\left(\delta_2 +\!\!\sqrt{-\frac{b}{c}}(x +\!\frac{b^2\text{v}}{c} t) \right)\right]\right\},
\end{equation}
where both
$\text{v}$ and $\delta_2$ are arbitrary.
Solution (\ref{usechsix7thKdVMinusorg}) satisfies (\ref{7thKdVorg}) provided $d = \frac{769\,c^2}{2500\,b}$.

\begin{figure}[hb!]
\centering
{\includegraphics[height=1.5in]{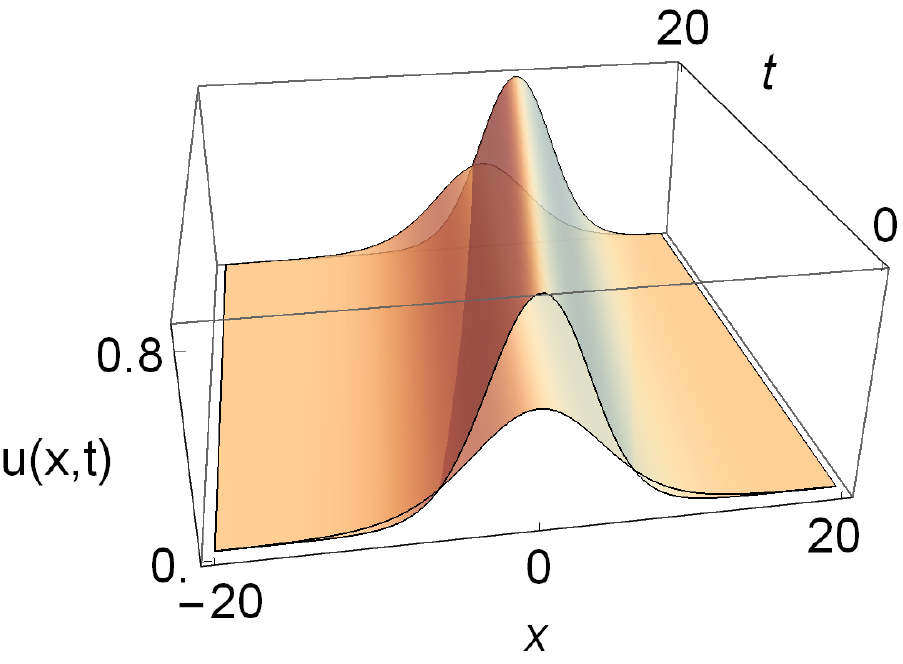}}
\caption{
Comparison of solutions (\ref{usechsix7thKdVPlusspec}) (short and wide)
and (\ref{usechsix7thKdVMinusspec}) (tall and narrow).
}
\label{fig:sech-7thkdv-comb}
\end{figure}

Setting $A_0 = 0$ fixes the wave speed $\text{v}^{-} = \frac{(18)\,10^4}{769^2} \approx 0.3044$ and $\mathcal{C}^{-} = 0$.
The solitary wave with speed $\text{v}^{-}$ approaches zero as $|x| \rightarrow \infty$.
Setting $\delta_2=0$ and replacing $\text{v}$ by $\text{v}^{-},$ further simplifies (\ref{usechsix7thKdVMinusorg})
into
\begin{equation}
\begin{array}{l}
\label{usechsix7thKdVMinusspec}
u(x,t) = -\,\frac{(154) 3^3 5^3\, b^2}{769^2\,ac}
\,\mathrm{sech}^6 \left[\frac{5}{\sqrt{1538}}\sqrt{-\frac{b}{c}}\left(x +\frac{(18)\,10^4\,b^2}{769^2\,c} t\right)\right].
\end{array}
\end{equation}
Fig.\ \ref{fig:sech-7thkdv-comb} shows solutions (\ref{usechsix7thKdVPlusspec}) and (\ref{usechsix7thKdVMinusspec})
for $a = b = - c = 1,$ resulting in $d = \frac{2159}{10^2}$ and $d = \frac{769}{2500},$ respectively.
Solitary wave (\ref{usechsix7thKdVMinusspec}) has approximately twice the speed,
twice the height, and three-quarters the width of solitary wave (\ref{usechsix7thKdVPlusspec}).
Setting $a=\frac{9}{10}$, $b=-\frac{11}{10}$, and $c=\frac{41}{40}$ would lead to a similar picture (not shown)
with two dips instead of two humps.
%
%
\section{Cnoidal Wave Solutions}
\label{cnsolutions}
To obtain cnoidal wave solutions we require that $Y_\xi = 0$ when $Y = 0$
which implies that $Q_3(Y)$ should have only one zero root, $Y_1 = 0,$
and two distinct real roots, $Y_2$ and $Y_3$.
This requires that $a_0 = 0.$
Under these assumptions (\ref{YxiEFM}) becomes
\begin{equation}
\label{Ucn}
{Y_\xi}^2 = a_1 Y + a_2 Y^2 + a_3 Y^3 = Y (a_1 + a_2 Y + a_3 Y^2),
\end{equation}
which can be factored as
\begin{equation}
\label{Yxicn}
{Y_\xi}^2 = - a_3 Y (Y_2 - Y)(Y_3 + Y).
\end{equation}

The solution of (\ref{Yxicn}) is the cnoidal wave \cite{KdV}
\begin{equation}
\label{Ycnsq}
Y(\xi)= Y_2\,\mathrm{cn}^2\left[\frac{1}{2} \sqrt{-a_3 (Y_2 + Y_3)}(\xi-\xi_0); k \right],
\end{equation}
where $\mathrm{cn}(\theta; k)$ is the Jacobi elliptic function with modulus
$k = \sqrt{\frac{Y_2}{Y_2 + Y_3}}$ and $\xi_0$ an arbitrary constant.
By equating the coefficients of (\ref{Ucn}) with (\ref{Yxicn}) we identify the roots,
$Y_{2} = - \frac{a_2 + \sqrt{\Omega}}{2 a_3}, \, Y_{3} = \frac{a_2 - \sqrt{\Omega}}{2 a_3},$
with discriminant and modulus given by
\begin{equation}
\label{discrmodulus}
\Omega = {a_2}^2 - 4 a_1 a_3 > 0, \quad k  = \sqrt{\frac{1}{2} \left( 1 + \frac{a_2}{\sqrt{\Omega}}\right)}.
\end{equation}
Thus, the solution of (\ref{Ucn}) becomes
\begin{equation}
\label{Ycnsqbis}
Y(\xi) =
  - \frac{1}{2} \left( \frac{a_2 + \sqrt{\Omega}}{a_3} \right)
    \,\mathrm{cn}^2\left[\frac{1}{2} \sqrt[4]{\Omega}(\xi-\xi_0);
    \sqrt{\frac{1}{2} \left( 1 + \frac{a_2}{\sqrt{\Omega}}\right)}\right], \;\; a_3 \ne 0.
\end{equation}
The expressions of the derivatives of $U$, after substitution of ${Y_\xi}^2$ from (\ref{Ucn})
together with all higher order derivatives, are given in (\ref{app:cnUderiv}).
Substituting these derivatives into (\ref{conslaw1}) results in a sixth-degree polynomial
$ \sum_{i=0}^6 s_i Y(\xi)^i, $
which must identically vanish since the powers of $Y(\xi)$ are independent for varying $\xi$.
The expressions for the coefficients $s_i$ are given in (\ref{app:cnscoef}).
The equations, $s_i = 0,$ are then solved for the $A_i$ in terms of the $a_i, \alpha,$ and $\text{v},$
some of which become constrained.
The integration constant ${\mathcal{C}}$ depends on $\text{v}$.
Instead of doing this in full generality, we consider special cases corresponding
to PDEs of physical relevance.
\subsection{Cn solutions for the fifth-order KdV equation}
\label{cnsolfifthorderKdV}
\begin{figure}[h!]
\centering
\includegraphics[height=1.5in]{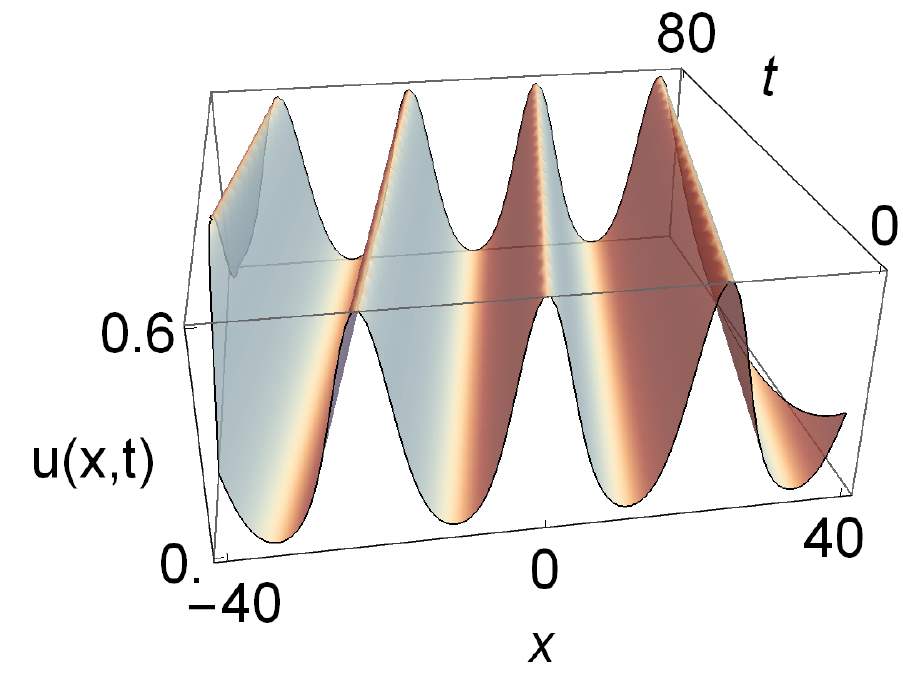}
\includegraphics[height=1.5in]{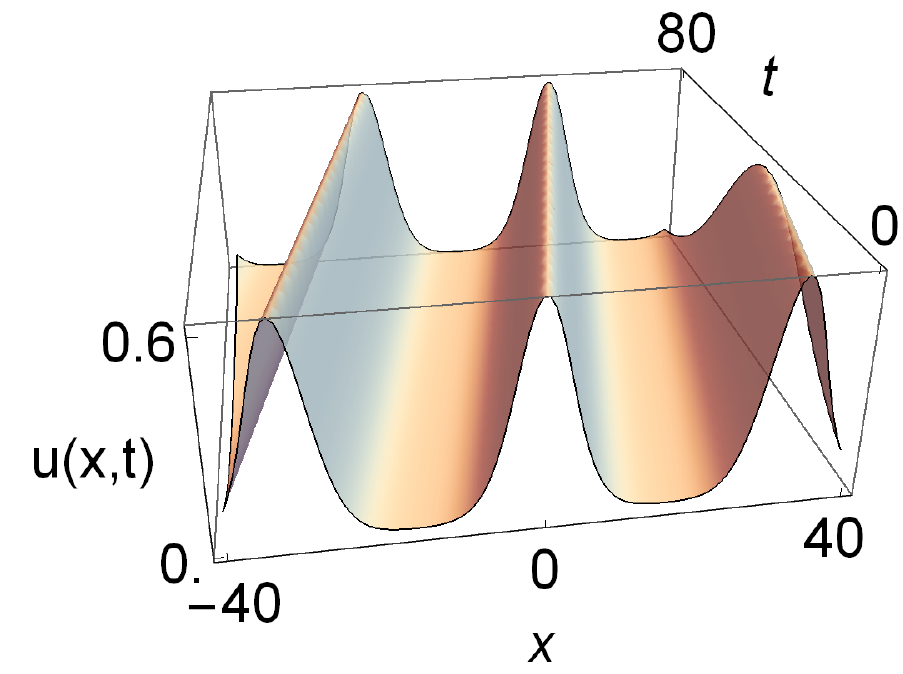}
\includegraphics[height=1.5in]{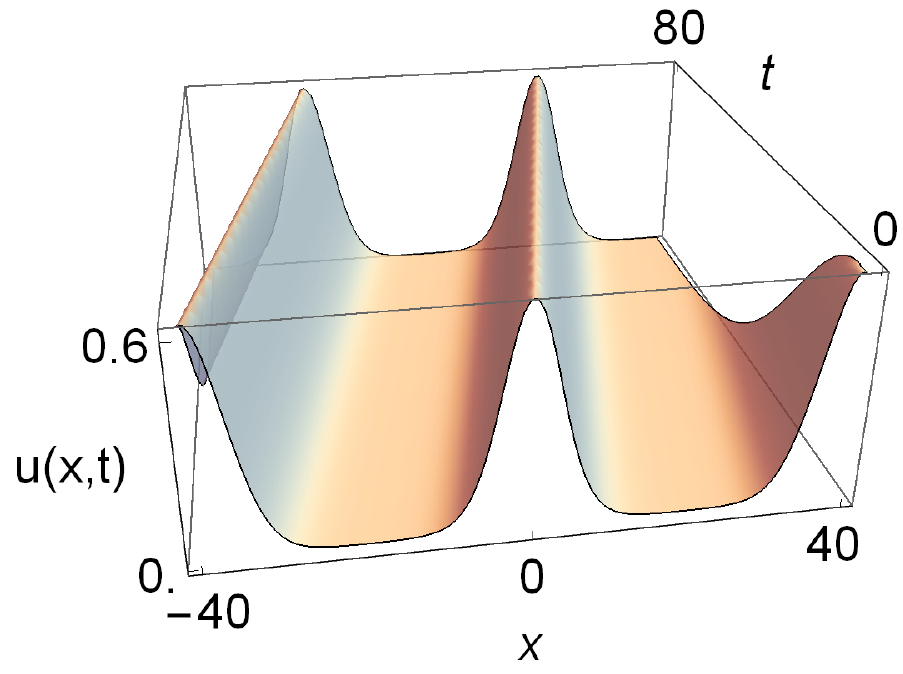}
\caption{
Solution (\ref{ucnquart5thKdVorg}) for $a = b = -c = 1$, \,
(a)\, $p = 0,$
(b)\, $p = \frac{3}{4},$
and (c)\, $p = \frac{9}{10}.$}
\label{fig:cn-5thkdv}
\end{figure}

We now compute cn solutions for (\ref{7thKdVXT}) with $\alpha = 0$.
Since $N=2$ in (\ref{UEFM}), we set $A_3 = \alpha = 0$ in (\ref{app:cnscoef}), discarding $s_5$ and $s_6.$
Solving $s_2 = s_3 = s_4 = 0$ yields the expansion coefficients
\begin{equation}
\label{cAcn5thKdV}
A_0 = \text{v} - \frac{31}{507} + 42 a_1 a_3 - \frac{7}{507} p (10 + p), \;\;
A_1 = -\frac{70}{13} a_3 (1 - p), \;\;
A_2 = 105 {a_3}^2,
\end{equation}
where $p = 13 a_2$.
Solving $s_1 = s_0 = 0,$ after inserting these coefficients, gives
\begin{equation}
\label{ca1cCcn5thKdV}
\begin{array}{ll}
& a_1 = -\, \frac{(1 - p)\left( 31 + 31 p + 10 p^2 \right)}{117 (13 a_3) \left(7 + 5 p \right)},
\\
& \mathcal{C} =
  \frac{7 \left[(63)13^4 \text{v}^2 - 124(677)\right]
     + 5p \left\{
     42 \left[(3)13^4 \text{v}^2 - 3658\right]
     - 5p \left[ 9(31)(47) + 161p (2-p) \left(2 -2 p -p^2 \right) - (9)13^4 \text{v}^2 \right]
  \right\}}{(18)13^4 (7 + 5 p)^2}.
\end{array}
\end{equation}
Substituting $a_1$ from (\ref{ca1cCcn5thKdV}) into (\ref{discrmodulus}) gives
\begin{equation}
\label{discrmoduluscn5thKdV}
\Omega = \frac{(2+p)\left(62 - 31 p + 5 p^2 \right) }{3^2 13^2 (7 + 5 p)},
\;\;
k = \frac{\sqrt{2}}{2} \sqrt{1 + 3p \sqrt{\frac{7+5p}{(2 + p)(62 - 31 p + 5 p^2)}}}.
\end{equation}
Finally, substituting $a_1$ into (\ref{cAcn5thKdV}), and these coefficients with (\ref{Ycnsq})
into (\ref{UEFM}), yields
\begin{eqnarray}
\label{Ucnquart5thKdV}
U(\xi) & = &
- \, \frac{7(31- 13^2\text{v})+ 5 p(43-13^2\text{v}) + 35 p^2(1-p)}{13^2 (7+5p)}
\nonumber \\
&& + \, \frac{35}{676}( p + 13 \sqrt{\Omega})
    \,\mathrm{cn}^2(\eta; k) \big[ 4(1-p)+ 3(p+13\sqrt{\Omega})\,\mathrm{cn}^2(\eta; k) \big],
\end{eqnarray}
where $\eta = \frac{1}{2} \sqrt[4]\Omega(\xi-\xi_0)$
with $\Omega$ and $k$ in (\ref{discrmoduluscn5thKdV}).
Note that for $p = 1$, (\ref{discrmoduluscn5thKdV}) yields $\Omega=\frac{1}{169}, k = 1,$
and the cnoidal wave (\ref{Ucnquart5thKdV}) becomes the solitary wave (\ref{Usechquart5thKdV}).
Converting (\ref{Ucnquart5thKdV}) into the original variables gives
\begin{eqnarray}
\label{ucnquart5thKdVorg}
u(x,t) & = &
\frac{b^2}{ac} \left\{ \frac{7(31 - 13^2\text{v}) + 5 p(43-13^2\text{v}) + 35 p^2(1-p)}{13^2 (7+5p)} \right.
\nonumber \\
&& \left.
- \frac{35}{676}( p + 13 \sqrt{\Omega})
\,\mathrm{cn}^2(\zeta; k)\big[ 4(1-p) + 3(p+13\sqrt{\Omega})\,\mathrm{cn}^2(\zeta; k) \big]
\right\},
\end{eqnarray}
where $\zeta =
\frac{1}{2} \sqrt[4]\Omega \left[ \delta_1 + \sqrt{-\frac{b}{c}}\left(x+\frac{b^2\text{v}}{c} t\right) \right]$,
satisfies (\ref{7thKdVorg}) with $d = 0$.
Eq.\ (\ref{ucnquart5thKdVorg}) expresses a new two-parameter family of solutions since
both $\text{v}$ and $p$ are arbitrary.
Only the solution with $\text{v} = \frac{128}{507} \approx 0.2525$ and $p = 0$ yielding $k = \frac{\sqrt{2}}{2}$
has been reported in the literature \cite{Manc}.
As before, we plot the solutions corresponding to $A_0=0$ yielding
$\text{v} = \frac{217+5p\left[43+7p(1-p)\right]}{169(7+5p)}$.
Fig.\ \ref{fig:cn-5thkdv} shows solution (\ref{ucnquart5thKdVorg}) for $a=b=-c=1$, with
$ p = 0,\, \frac{3}{4},$ and $\frac{9}{10}$
for which $\text{v} = \frac{31}{169} \approx 0.1834,\, \frac{137 (179)}{(43)\,2^4\,13^2} \approx 0.2109$,
and $\frac{6359}{(299)\,10^2} \approx 0.2127,$
respectively.
The cnoidal waves spread out more as $p$ increases.
As $p$ approaches one the cnoidal wave turns into solitary wave (\ref{Usechquart5thKdV}) with speed
$\text{v}=\frac{36}{169} \approx 0.2130$.
\subsection{Cn solutions for the seventh-order KdV equation}
\label{cnsolseventhorderKdV}
To find cnoidal wave solutions for (\ref{7thKdVXT}) using the expressions in (\ref{app:cnscoef}),
we first solve $s_3 = s_4 = s_5 = s_6 = 0$, yielding
\begin{align}
\label{cAcn7thKdV}
& A_0 = \text{v}
  - \frac{231}{50} {a_2}^2 -22 \alpha {a_2}^3  + \frac{33}{25} q (7 - 50 \alpha a_2)
  + \frac{(334)7^2 - (681)5^3 \alpha}{(262)5^5 \alpha^2} + \frac{231 (49 - 500 \alpha) a_2}{(262)5^3 \alpha},
\nonumber \\
& A_1 = \frac{(77) 3^2 a_3}{(262) 5^3 \alpha} \left[ 49 - 50 \alpha  \left(10 - 131 a_2 \right)
         - (262) 5^3 \alpha^2 \left( 2 {a_2}^2 + q \right) \right],
\\
& A_2 = \frac{231}{10}(3 a_3)^2 \left(1 - 50 \alpha a_2 \right), \;\;
A_3 = - 385 \alpha (3 a_3)^3,
\nonumber
\end{align}
where $q = 9 a_1 a_3.$

Upon substitution of these coefficients into $s_2 = 0,$ we obtain a complicated equation
(not shown) of fourth degree in both $\alpha$ and $a_2$ and quadratic in $q$.
Likewise, $s_1 = 0$ yields an equation of fifth degree in both $\alpha$ and $a_2$ and quadratic in $q,$
and $s_0 = 0,$ which would determine ${\mathcal{C}}$, becomes of degree six in both $\alpha$ and $a_2$
and cubic in $q$.
Inspired by the results in Sect.~\ref{cnsolfifthorderKdV}, we illustrate the solution procedure
for $a_2 = 0.$
The numerical treatment of the general case $(a_2 \ne 0)$ would be similar but very cumbersome.

For $a_2 =0,$ from $s_1 = 0$ and $s_2 = 0$ we get
\begin{align}
\label{s1s2cn7thKdV}
& 6 (167) 7^4 -\! 125 \alpha \left\{3 (2017) 7^2 - \alpha \left[(227) 6^2 \, 5^3 -\!
       77 {\tilde{q}} \left( 441 -\! 500 \alpha (9 - 4 {\tilde{q}} \alpha) \right)\right]
       \right\} = 0, \nonumber \\
& 3 (37)(2671) 7^2 \!-\! 500 \alpha \left\{ 6 (101) 317 \!-\! \alpha \left[6 (331) 5^3 \!+\!
       77 {\tilde{q}} \left(147 \!-\! 250 \alpha (6 \!-\! {\tilde{q}} \alpha)\right)\right]
       \right\} = 0,
\end{align}
respectively, where ${\tilde{q}} = 131 q = 9 (131) a_1 a_3.$

Solving (\ref{s1s2cn7thKdV}) numerically yields the following four real solutions for the couples
$({\tilde{q}}, \alpha)$:
$(-2.4191, 0.3075),\, (-1.2262, 0.2158),\, (2.2502, 1.6115)$, and $(5.6486, 0.1912)$.
Setting $a_2 = 0$ in (\ref{discrmodulus}) gives
$\Omega = - 4 a_1 a_3 = - \frac{4 {\tilde{q}} }{9 (131)} > 0$ and $ k = \frac{\sqrt{2}}{2},$
requiring ${\tilde{q}} < 0$.
Using $q = \frac{{\tilde{q}}}{131},$ only
\begin{equation}
\label{qialphaicn7thKdV}
(q_1, \alpha_1) = (-0.0185, 0.3075) \;\; {\rm and} \;\; (q_2, \alpha_2) = (-0.0094, 0.2158)
\end{equation}
will lead to real solutions.
We label the solutions below accordingly.

\begin{figure}[ht!]
\centering
\includegraphics[height=1.55in]{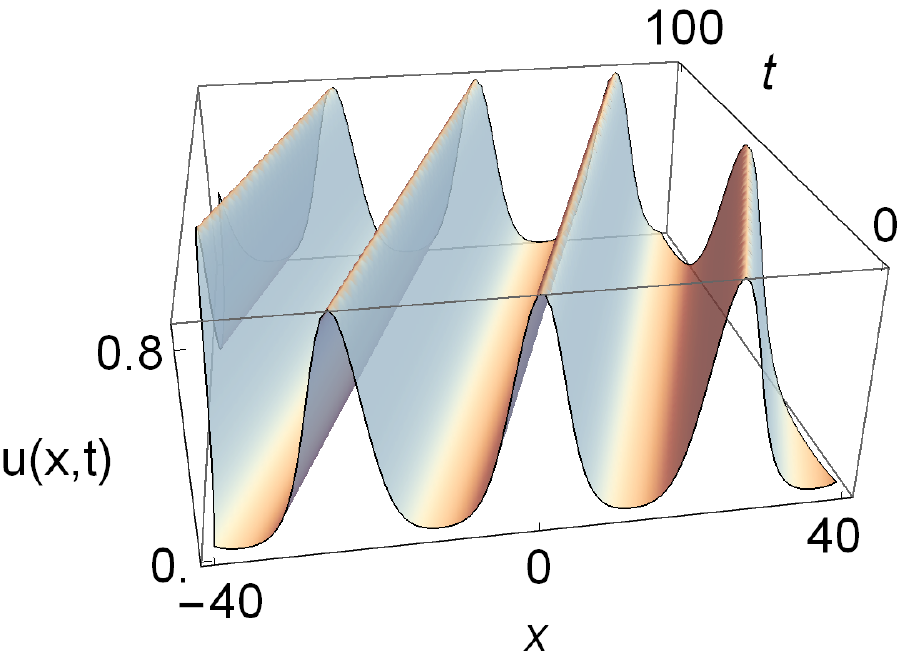}
\includegraphics[height=1.55in]{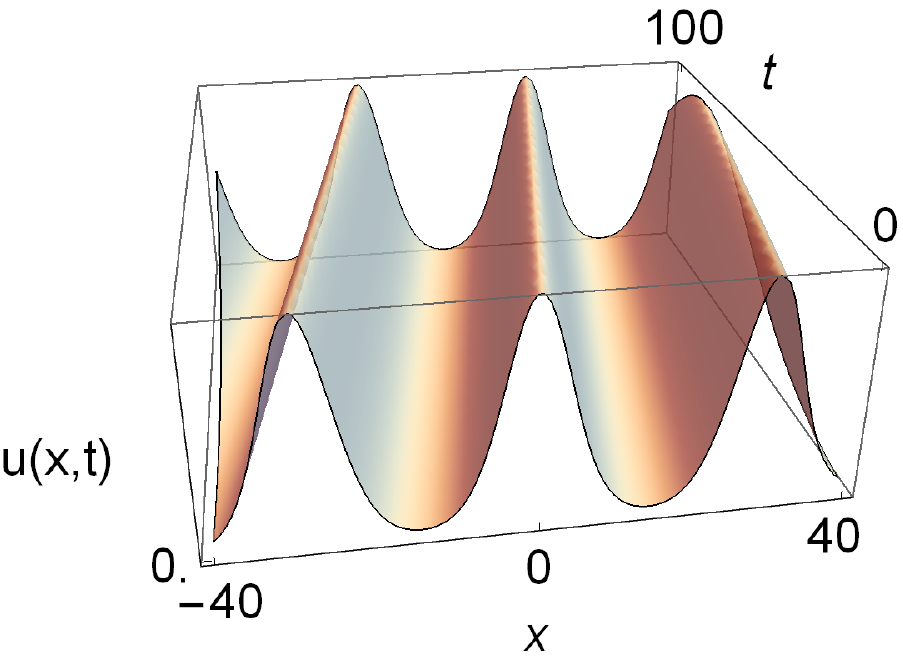}
\caption{
Solution (\ref{ucnsix7thKdVorg}) for $a = b = -c = 1$
for (a)\, $q_1=-0.0185, \, \alpha_1=0.3075, \, \text{v}_1=0.2973,$
and (b)\, $q_2=-0.0094, \, \alpha_2=0.2158, \, \text{v}_2=0.1391.$}
\label{fig:cn-7thkdv}
\end{figure}

Finally, substituting (\ref{cAcn7thKdV}) with $a_2 = 0$ into (\ref{UEFM}) gives
two solutions
\begin{eqnarray}
\label{Ucnsix7thKdV}
U_i(\xi) & = &
{ \text{v} + \frac{231}{25} q_i - \frac{681}{2 (131) 5^2 \alpha_i} + \frac{(167) 7^2}{(131) 5^5 \alpha_i^2} }
\nonumber \\
&& +\, { \frac{(77) 3^2 a_3}{(262) 5^3 \alpha_i} \Big( 49-500 \alpha_i -(262) 5^3 q_i {\alpha_i}^2 \Big) } Y_i(\xi)
\nonumber \\
&& +\, { \frac{231}{10}(3 a_3)^2 } Y_i(\xi)^2 - { 385 \alpha_i (3 a_3)^3 } Y_i(\xi)^3,
\end{eqnarray}
where
\begin{equation}
\label{Ycnsq7thKdV}
Y_i(\xi) = -\, \frac{\sqrt{-q_i}}{3 a_3}
    \,\mathrm{cn}^2 \left[ \frac{\sqrt[4]{-q_i}}{\sqrt{6}}\Big(\xi-\xi_0\Big);
    \frac{\sqrt{2}}{2} \right], \; q_i < 0, \; i = 1, 2.
\end{equation}
In the original variables,
\begin{eqnarray}
\label{ucnsix7thKdVorg}
u_i(x,t) & = &  -\, \frac{b^2}{ac}
\left\{ \text{v} + \frac{231}{25} q_i - \frac{681}{2 (131) 5^2 \alpha_i} + \frac{(167) 7^2}{(131) 5^5 \alpha_i^2} \right.
\nonumber \\
&& \left. -\, { \frac{(231) \sqrt{-q_i}}{(262) 5^3 \alpha_i} \Big( 49-500 \alpha_i -(262) 5^3 q_i {\alpha_i}^2 \Big) }
\,\mathrm{cn}^2\left[\zeta_i; \frac{\sqrt{2}}{2}\right] \right.
\nonumber \\
&& \left. -\, { \frac{231 q_i}{10}}\,\mathrm{cn}^4\left[\zeta_i; \frac{\sqrt{2}}{2}\right]
          -\, { 385 \alpha_i q_i \sqrt{-q_i} }\,\mathrm{cn}^6\left[\zeta_i; \frac{\sqrt{2}}{2}\right] \right\},
          \; q_i < 0, \; i = 1, 2,
\end{eqnarray}
where
$\zeta_i =
\frac{\sqrt[4]{-q_i}}{\sqrt{6}} \left[ \delta_i + \sqrt{-\frac{b}{c}}\left(x+\frac{b^2\text{v}}{c} t\right) \right]$,
satisfying (\ref{7thKdVorg}) provided $d = d_i = \frac{c^2 \alpha_i}{b}, $ and with $(q_i, \alpha_i)$
given in (\ref{qialphaicn7thKdV}).
Ignoring the phase constant $\delta_i$, each solution in (\ref{ucnsix7thKdVorg}) defines
a one-parameter family of solutions since $\text{v}$ is arbitrary.
To our knowledge, solutions of type (\ref{ucnsix7thKdVorg}) have not been reported in the literature.
To remove the constant term in (\ref{ucnsix7thKdVorg}), we set
$\text{v}_i = \frac{375 \alpha_i \left[ 227 - 131(154) q_i\alpha_i \right]- (334)\,7^2}{(262)\,5^5 \alpha_i^2}$,
and from $(\ref{qialphaicn7thKdV})$ we then have $\text{v}_1 \approx 0.2973$ and $\text{v}_2 \approx 0.1391$, respectively.
These solutions are presented in Fig.\ \ref{fig:cn-7thkdv} for $a=b=-c=1$.
\section{Discussion and Conclusions}
\label{concl}
In this paper we surveyed the origin and applications of a KdV equation with higher-order dispersive terms.
Specific members of this parameterized family of equations describe shallow water waves, electrical pulses
in transmission lines, magneto-acoustic and hydrodynamic waves in plasmas, etc.
Using an elliptic function method, we computed hump-type solitary waves and cnoidal wave solutions
which can be used in bifurcation analyses and as benchmarks for both numerical solvers and perturbation methods.

Some of the solutions presented in this paper correct, complement, and illustrate results previously
reported in the literature.
Solution (\ref{usechsix7thKdVPlusorg}) was missed by Ma \cite{Ma} but later computed
by Duffy and Parkes \cite{Duffy}.
Solution (\ref{usechsix7thKdVMinusorg}) is equivalent to a solution reported by Ma \cite{Ma}.
However, solution (6) on p.\ 222 in Ma's paper \cite{Ma} should have read
\begin{equation}
\label{usechsix7thKdVMinusspecMa}
u(x,t) = - \frac{519750\,b^2}{769^2\,ac}
\,\mathrm{sech}^6 \left\{ \sqrt{-\frac{25\,b}{1538\,c}}
\left[x - \left(a E - \frac{180000\,b^2}{769^2\,c} \right) t - \xi_0 \right] \right\} + E,
\end{equation}
where $E$ and $\xi_0$ are arbitrary constants.
To convert (\ref{usechsix7thKdVMinusspecMa}) into (\ref{usechsix7thKdVMinusorg}),
set $E = \frac{b^2}{769^2\,ac} \left(180000-769^2\, \text{v} \right)$
and $\delta_2 = \sqrt{-\frac{b}{c}} \,\xi_0.$

Solutions (\ref{usechsix7thKdVPlusorg}) and (\ref{usechsix7thKdVMinusorg}) to (\ref{7thKdVorg})
are the only polynomial solutions involving the sech-function.
Both require that the coefficients in (\ref{7thKdVorg}) satisfy specific algebraic relations.
Application of the tanh- or sech-method \cite{Bald} to (\ref{7thKdVXT}) results in expressions
which are equivalent to (\ref{Usechsix7thKdVPlus}) and (\ref{Usechsix7thKdVMinus}) as verified with the
{\it Mathematica} package {\sc PDESpecialSolutions.m} \cite{BaldSoft}.
To our knowledge, solutions (\ref{ucnquart5thKdVorg}) and (\ref{ucnsix7thKdVorg}) are novel
although a special case of (\ref{ucnquart5thKdVorg}) had been reported \cite{Manc}.
Solutions as complicated as (\ref{ucnquart5thKdVorg}) and (\ref{ucnsix7thKdVorg})
are beyond the present capabilities of {\sc PDESpecialSolutions.m}.

Finally, all solutions reported in this paper have been verified with {\it Mathematica}
which uses the {\it square} of the modulus (denoted by $k$ in this paper) of the Jacobi elliptic functions.
\vfill
\newpage
\section*{Figures}
%
%
%
%
\phantom{.}
\vspace*{-4mm}
%
%
%
%
\vfill
\vspace*{-4mm}
\phantom{.}
\vspace*{-4mm}

\vfill
\newpage
\appendix
\numberwithin{equation}{section}
\section{}
\label{appendixA}
\subsection{Formulas needed to compute sech solutions}
\label{sechformulas}
\begin{align}
\label{app:sechUderiv}
& U_{2\xi} =
\frac{Y}{2} \left[ 2A_1a_2 + \left(8A_2a_2+3 A_1a_3\right)Y + 2\left(5 A_2a_3+9A_3 a_2\right)Y^2 + 21 A_3a_3 Y^3 \right],
\nonumber \\
& U_{4\xi} = \frac{Y}{2}
\left[2A_1{a_2}^2  +  a_2\left(32A_2a_2  +  15 A_1a_3\right)Y
 +  \left(130 A_2a_2a_3  +  15A_1 {a_3}^2  +  162 A_3{a_2}^2\right)Y^2 \right.
\nonumber \\
& \left. \;\;\;\;\;\;\;\;\; + 105a_3\left(A_2a_3+5A_3a_2\right)Y^3 + 378A_3{a_3}^2Y^4 \right],
\\
& U_{6\xi} = \frac{Y}{4}
\left[ 4A_1{a_2}^3+2{a_2}^2\left(128A_2a_2+63 A_1a_3 \right)Y + 4 a_2\left(665 A_2a_2a_3 + 105A_1 {a_3}^2 \right. \right.
\nonumber \\
& \left. \left. \;\;\;\;\;\;\;\;\;
+ A_3 {(27 a_2)}^2\right)Y^2 + 21 a_3 \left(290A_2a_2a_3+15 A_1{a_3}^2+962 A_3{a_2}^2\right)Y^3 \right.
\nonumber \\
& \left. \;\;\;\;\;\;\;\;\; + 105 (6 a_3)^2 \left(A_2a_3+10A_3a_2\right)Y^4 + 770 A_3 (3 a_3)^3 Y^5 \right].
\nonumber
\end{align}

\begin{align}
\label{app:sechrcoef}
& r_0 = \frac{A_0}{2} (2 \text{v} - A_0) - \mathcal{C},
\nonumber \\
& r_1 = A_1 \left[\text{v} - A_0 - a_2 (1 - a_2 + \alpha {a_2}^2) \right],
\nonumber \\
& r_2 = - \frac{A_1}{2} \left[ A_1 + 3 a_3 \left(1 - 5 a_2 + 21 \alpha {a_2}^2 \right) \right]
        + A_2 \left[ \text{v} - A_0
        - 4 a_2 \left(1 - 4 a_2 + 16 \alpha {a_2}^2 \right) \right],
\nonumber \\
& r_3 =  A_1 \left[\frac{15}{2} {a_3}^2 (1 - 14 \alpha a_2) - A_2 \right]
        - 5 A_2 a_3 \left( 1 - 13 a_2 + 133 \alpha {a_2}^2 \right)
\nonumber \\
& \;\;\;\;\;\;\; + A_3 \left[ \text{v} -A_0 - 9 a_2 \left(1 - 9 a_2 +81\alpha {a_2}^2 \right) \right],
\\
& r_4 = - A_1 \left(A_3 + \frac{315}{4} \alpha {a_3}^3 \right)
        - \frac{{A_2}}{2} \left[ A_2 - 105 {a_3}^2 \left(1 - 29 \alpha a_2 \right) \right]
\nonumber \\
& \;\;\;\;\;\;\; - \frac{21}{2} A_3 a_3 \left( 1 - 25 a_2 + 481 \alpha {a_2}^2 \right),
\nonumber \\
& r_5 = - A_2 \left[ A_3 + 35 \alpha {(3 a_3)}^3 \right] + 21 A_3 {(3 a_3)}^2 \left(1 - 50 \alpha a_2 \right),
\nonumber \\
& r_6 = -\frac{A_3}{2} \left[ A_3 + 385 \alpha (3 a_3)^3 \right].
\nonumber
\end{align}

\subsection{Formulas needed to compute cn solutions}
\label{cnformulas}

\begin{align}
\label{app:cnUderiv}
& U_{2\xi} = \frac{1}{2} \left[A_1a_1 + 2\left(A_1a_2+3 A_2a_1\right)Y + \left(3A_1 a_3+8 A_2a_2+15 A_3a_1\right)Y^2 \right.
\nonumber \\
& \left. \;\;\;\;\;\;\;\;\;+ 2\left(5 A_2a_3+9 A_3a_2\right) Y^3 +21 A_3a_3 Y^4\right],
\nonumber \\
& U_{4\xi} = \frac{1}{2}
\left\{a_1\left(A_1a_2+3A_2a_1\right)  +  \left[A_1\left(2{a_2}^2+9a_1a_3\right)+30 A_2a_1a_2+45 A_3{a_1}^2\right]Y\right.
\nonumber \\
& \left. \;\;\;\;\;\;\;\;\;
 +  \left[15a_2(A_1a_3+13 A_3a_1)+ 4 A_2( 8 {a_2}^2 + 21 a_1 a_3) \right]Y^2 \right.
\nonumber \\
& \left. \;\;\;\;\;\;\;\;\; + \left[ 5 a_3 (3 A_1 a_3 + 26 A_2 a_2) + 9 A_3 (18 a_2^2 + 41 a_1 a_3) \right] Y^3 \right.
\nonumber \\
& \left. \;\;\;\;\;\;\;\;\;+105{a_3}(A_2a_3+5A_3a_2)Y^4+378A_3{a_3}^2Y^5 \right\},
\nonumber \\
& U_{6\xi} = \frac{1}{4}
\left\{ a_1\left[2A_1{a_2}^2+3a_1(3 A_1a_3+10A_2a_2+15A_3a_1)\right] \right.
\\
& \left. \;\;\;\;\;\;\;\;\;+4\left[A_1 a_2 \left(27 a_1 a_3+{a_2}^2\right)+63 a_1\left(A_2(2a_1a_3+{a_2}^2)
         +5 A_3a_1a_2\right) \right]Y \right.
\nonumber \\
& \left. \;\;\;\;\;\;\;\;\;
 + 2\left[4 A_2a_2\left(339 a_1a_3+32 {a_2}^2\right)+63a_1a_3(2 A_1 a_3+45 A_3a_1)  \right. \right.
\nonumber \\
& \left. \left. \;\;\;\;\;\;\;\;\; + 21{a_2}^2(3 A_1a_3+95A_3a_1)\right]Y^2  \right.
\nonumber \\
& \left. \;\;\;\;\;\;\;\;\;+ 4\left[35 A_2a_3\left(27 a_1a_3+19 {a_2}^2\right)+ 729 A_3{a_2}^3
          +3 a_2a_3(35 A_1 a_3+1941 A_3a_1)\right]Y^3 \right.
\nonumber \\
& \left. \;\;\;\;\;\;\;\;\; + 21a_3 \left[2a_2(145A_2a_3+481A_3a_2)+3a_3(5 A_1a_3+393 A_3a_1)\right]Y^4  \right.
\nonumber \\
& \left. \;\;\;\;\;\;\;\;\; + 105 {(6 a_3)}^2 (A_2a_3+10A_3a_2) Y^5+770A_3(3a_3)^3Y^6 \right\}.
\nonumber
\end{align}

\begin{align}
\label{app:cnscoef}
& s_0 =\frac{A_0}{2} (2 \text{v} - A_0) - \mathcal{C}
\nonumber \\
& \;\;\;\;\;\;\;\;\; - \frac{a_1}{2} \left[ A_1 \left(1-a_2 + \alpha{a_2}^2 + \frac{9}{2} \alpha a_1 a_3  \right)
                     - 3 A_2 a_1\left( 1 - 5 \alpha a_2 \right)  + \frac{45}{2} A_3 \alpha {a_1}^2 \right],
\nonumber \\
& s_1 = A_1 \left[\text{v}-A_0 - a_2 (1-a_2 +\alpha {a_2}^2 +27 \alpha a_1 a_3) + \frac{9}{2} a_1 a_3 \right]
\nonumber \\
&  \;\;\;\;\;\;\;\;\; - 3 a_1 \left[ A_2 \left(1 -5a_2 + 21\alpha ({a_2}^2 +2 a_1 a_3) \right)
                      -\frac{15}{2} A_3 a_1(1-14 \alpha a_2)\right],
\nonumber \\
& s_2 = - \frac{A_1}{2} \left[ A_1 + 3 a_3 \left(1-5a_2+21\alpha({a_2}^2+2 a_1a_3)\right)\right]
\nonumber \\
& \;\;\;\;\;\;\;\;\; + A_2 \left[\text{v} - A_0   - 4a_2( 1-4 a_2 + 16\alpha {a_2}^2) +6 a_1 a_3 (7-113\alpha a_2) \right]
\nonumber \\
& \;\;\;\;\;\;\;\;\; - \frac{15}{2} A_3 a_1 \left[ 1 - a_2(13 -133 \alpha a_2) +189 \alpha a_1a_3 \right],
\\
& s_3 = A_1 \left[ \frac{15}{2} {a_3}^2 (1-14 \alpha a_2) - A_2 \right]
        -5 A_2 a_3\left[1-13 a_2 + 7 \alpha (19 {a_2}^2 + 27 a_1 a_3)\right]
\nonumber \\
& \;\;\;\;\;\;\;\;\; + A_3\left[\text{v} -A_0 - 9 a_2 \left(1 - 9 a_2 +81\alpha {a_2}^2 + 647 \alpha a_1 a_3 \right)
                     +\frac{369}{2} a_1 a_3 \right],
\nonumber \\
& s_4 = -A_1 (A_3 + \frac{315}{4} \alpha {a_3}^3) - \frac{A_2}{2} \left[ A_2 - 105 {a_3}^2 (1-29 \alpha a_2)\right]
\nonumber \\
& \;\;\;\;\;\;\; -\frac{21}{2} A_3 a_3 \left[1 -25 a_2 +481 \alpha {a_2}^2 +\frac{1179}{2} \alpha a_1 a_3\right],
\nonumber \\
& s_5 = - A_2 \left[A_3 + 35 \alpha {(3 a_3)}^3 \right] + 21 A_3 {(3 a_3)}^2 \left(1 - 50 \alpha a_2 \right),
\nonumber \\
& s_6 = - \frac{A_3}{2} \left[A_3 + 385 \alpha (3 a_3)^3 \right].
\nonumber
\end{align}
%
%


\newpage

\begin{thebibliography}{99}

\bibitem{Bald}
D.\ Baldwin, \"{U}.\ G\"{o}kta\c{s}, W.\ Hereman, L.\ Hong, R.\ S.\ Martino, and J.\ C.\ Miller,
J.\ Symb.\ Comp.\ {\bf 37}, 669 (2004).

\bibitem{Malf1}
W.\ Malfliet,
Am.\ J.\ Phys.\ {\bf 60}, 650 (1992).

\bibitem{Malf2}
W.\ Malfliet and W.\ Hereman,
Phys.\ Scr.\ {\bf 54}, 563 (1996).

\bibitem{Malf3}
W.\ Malfliet and W.\ Hereman,
Phys.\ Scr.\ {\bf 54}, 569 (1996).

\bibitem{KdV}
D.\ J.\ Korteweg and G.\ de Vries,
Philos.\ Mag.\ (Ser.\ 5) {\bf 39}, 422 (1895).

\bibitem{Has}
H.\ Hasimoto,
Kagaku Sci.\ {\bf 40}, 401 (1970) [In Japanese].

\bibitem{Kawa5}
T.\ Kawahara,
J.\ Phys.\ Soc.\ Jpn.\ {\bf 33}, 260 (1972).

\bibitem{Kaku}
T.\ Kakutani and H.\ Ono,
J.\ Phys.\ Soc.\ Jpn.\ {\bf 26}, 1305 (1969).

\bibitem{Craig}
W.\ Craig, P.\ Guyenne, and H.\ Kalisch,
Commun.\ Pure Appl.\ Math.\ {\bf 58}, 1587 (2005).

\bibitem{Pom}
Y.\ Pomeau, A.\ Ramani, and B.\ Grammaticos,
Physica D {\bf 31}, 127 (1988).

\bibitem{Naga1}
H.\ Nagashima,
J.\ Phys.\ Soc.\ Jpn.\ {\bf 47}, 1387 (1979).

\bibitem{Ros1}
P.\ Rosenau,
Phys.\ Scr.\ {\bf 34}, 827 (1986).

\bibitem{Ros2}
P.\ Rosenau,
Prog.\ Theor.\ Phys.\ {\bf 79}, 1028 (1988).

\bibitem{Ma}
W.-X.\ Ma,
Phys.\ Lett.\ A {\bf 180}, 221 (1993).

\bibitem{Duffy}
B.\ R.\ Duffy and E.\ J.\ Parkes,
Phys.\ Lett.\ A {\bf 214}, 271 (1996).

\bibitem {Parkes}
E.\ J.\ Parkes, Z.\ Zhu, B.\ R.\ Duffy, and H.\ C.\ Huang,
Phys.\ Lett.\ A {\bf 248}, 219 (1998).

\bibitem{Yang}
Z.\ J.\ Yang,
Int.\ J.\ Theor.\ Phys.\ {\bf 34}, 641 (1995).

\bibitem{Dai}
X.\ Dai and J.\ Dai,
Phys.\ Lett.\ A {\bf 142}, 367 (1989).

\bibitem{Hunter83}
J.\ K.\ Hunter and J.-M.\ Vanden-Broeck,
J.\ Fluid Mech.\ {\bf 134}, 205 (1983).

\bibitem{Her85}
W.\ Hereman, A. Korpel, and P.\ P.\ Banerjee,
Wave Motion {\bf 7}, 283 (1985).

\bibitem{Her86}
W.\ Hereman, P.\ P.\ Banerjee, A.\ Korpel, G.\ Assanto, A.\ Van Immerzeele, and A.\ Meerpoel,
J.\ Phys.\ A {\bf 19}, 607 (1986).

\bibitem{Kud3}
N.\ A.\ Kudryashov,
J.\ Comput.\ Appl.\ Math.\ {\bf 234}, 3511 (2010).

\bibitem{Manc}
S.\ C.\ Mancas,
to be published in Differ.\ Equ.\ Dyn.\ Syst.\ DOI 10.1007/s12591-017-0367-5 (2017).

\bibitem{Andrade}
T.\ P.\ de Andrade, F.\ Crist\'{o}fani, and F.\ Natali,
J.\ Math.\ Phys.\ {\bf 58}, 051504 (2017).

\bibitem{Yama7}
Y.\ Yamamoto and \'{E}.\ I.\ Takizawa,
J.\ Phys.\ Soc.\ Jpn.\ {\bf 50}, 1421 (1981).

\bibitem{Kano4}
K.\ Kano and T.\ Nakayama,
J.\ Phys.\ Soc.\ Jpn.\ {\bf  50}, 361 (1981).

\bibitem{Khater}
A.\ H.\ Khater, M.\ M.\ Hassan and R.\ S.\ Temsah,
Math.\ Comput.\ Simulat.\ {\bf 70}, 221 (2005).

\bibitem{Huang}
G.-X.\ Huang, S.-Y.\ Luo, and X.-X. Dai,
Phys.\ Lett.\ A {\bf 139}, 373 (1989).

\bibitem{Parkesbis}
E.\ J.\ Parkes, B.\ R.\ Duffy, and P.\ C.\ Abbott,
Phys.\ Lett.\ A {\bf 295}, 280 (2002).

\bibitem{Kalischetal}
H.\ Kalisch, D.\ Moldabayev, and O.\ Verdier,
Electron.\ J.\ Differ.\ Eq.\ {\bf 2017}, 1 (2017).

\bibitem{Simpson}
G.\ Simpson and M.\ Spiegelman,
J.\ Sci.\ Comput.\ {\bf 49}, 268 (2011).

\bibitem{Longetal}
Y.\ Long, W.\ Rui, and B.\ He,
Chaos Solitons Fract.\ {\bf 23}, 469 (2005).

\bibitem{Lietal}
J.\ Li, X.\ Li, and W.\ Zhang,
J.\ Appl.\ Anal.\ Comput.\ {\bf 7}, 841 (2017).

\bibitem{Arora}
R.\ Arora and H.\ Sharma,
Int.\ J.\ Syst.\ Assur.\ Eng.\ Manag.\ {\bf 9}, 131 (2018).

\bibitem{Din}
S.\ T.\ Mohyud-Din, M.\ A.\ Noor, and K.\ I.\ Noor,
Int.\ J.\ Nonl.\ Sci.\ Numer.\ Simul.\ {\bf 10}, 227 (2009).

\bibitem{Dinbis}
S.\ T.\ Mohyud-Din, M.\ A.\ Noor, and K.\ I.\ Noor,
Int.\ J.\ Mod.\ Phys.\ B {\bf 23}, 3265 (2009).

\bibitem{Din2bis}
S.\ T.\ Mohyud-Din, M.\ A.\ Noor, and F.\ Jabeen,
Int.\ J.\ Comput.\ Meth.\ Engr.\ Sci.\ Mech.\ {\bf 12}, 107 (2011).

\bibitem{Drazin}
P.\ G.\ Drazin and R.\ S.\ Johnson,
{\em Solitons: An Introduction}
(Cambridge University Press, Cambridge, U.K., 1989).

\bibitem{BaldSoft}
D.\ Baldwin, \"{U}.\ G\"{o}kta\c{s}, W.\ Hereman, L.\ Hong, R.\ S.\ Martino, and J.\ C.\ and  Miller,
\verb|PDESpecialSolutions.m|, URL: \verb|http://inside.mines.edu/~whereman/| (2004).
\end{thebibliography}
\end{document}